\input psfig.tex
\input mn.tex


\def\fr#1#2{\textstyle {#1\over #2}\displaystyle}

\overfullrule=0pt
\def\d{{\rm d}} 
\def\bmu{\bar \mu}
\def\bnu{\bar \nu}
\def\vlos{\langle v_{\rm los} \rangle}
\def\vlossq{\langle v_{\rm los}^2 \rangle}

\def\piby2{{\pi \over 2}}
\def\threeh{{3\over 2}}

\def\xppdot{{\dot x}''}
\def\yppdot{{\dot y}''}
\def\zppdot{{\dot z}''}
\def\vlos{\langle v_{\rm los} \rangle}

\def\mux{\langle \mu_{x''}\rangle}
\def\muy{\langle \mu_{y''}\rangle}
\def\muz{\langle \mu_{z''}\rangle}
\def\muxx{\langle \mu^2_{x''x''}\rangle}
\def\muyy{\langle \mu^2_{y''y''}\rangle}
\def\muzz{\langle \mu^2_{z''z''}\rangle}

\def\vc{v_{\rm c}}
\def\vmin{{v_{\rm min}}}
\def\mN{{m_{\rm N}}}
\def\RN{{R_{\rm N}}}
\def\mr{{m_{\rm r}}}
\def\mX{{m_{\chi}}}
\def\ET{{E_{\rm T}}}
\def\cm{\,{\rm cm}}
\def\sigR{{\sigma_R}}
\def\sigz{{\sigma_z}}
\def\sigphi{{\sigma_\phi}}
\def\vearth{v_\oplus}
\def\fs{f_{\rm s}}
\def\tp{t_{\rm p}}
\def\kms{\,{\rm kms}^{-1}}
\def\gammamod{{\alpha}}

%
%
\def\spose#1{\hbox to 0pt{#1\hss}}
\def\lta{\mathrel{\spose{\lower 3pt\hbox{$\sim$}}
    \raise 2.0pt\hbox{$<$}}}
\def\gta{\mathrel{\spose{\lower 3pt\hbox{$\sim$}}
    \raise 2.0pt\hbox{$>$}}}
\def\today{\ifcase\month\or
 January\or February\or March\or April\or May\or June\or
 July\or August\or September\or October\or November\or December\fi
 \space\number\day, \number\year}
%
%
\newdimen\hssize
\hssize=8.4truecm  
\newdimen\hdsize
\hdsize=17.7truecm    


\newcount\eqnumber
\eqnumber=1
\def\chaphead{}
 
\def\new{\hbox{(\rm\chaphead\the\eqnumber)}\global\advance\eqnumber by 1}
 
\def\first{\hbox{(\rm\chaphead\the\eqnumber a)}\global\advance\eqnumber by 1}
\def\last#1{\advance\eqnumber by -1\hbox{(\rm\chaphead\the\eqnumber#1)}\advance
     \eqnumber by 1}
 
\def\ref#1{\advance\eqnumber by -#1 \chaphead\the\eqnumber
     \advance\eqnumber by #1}
 
\def\nref#1{\advance\eqnumber by -#1 \chaphead\the\eqnumber
     \advance\eqnumber by #1}

\def\eqnam#1{\xdef#1{\chaphead\the\eqnumber}}
 
 

\pageoffset{-0.85truecm}{-1.05truecm}



\pagerange{}
\pubyear{1996}
\volume{000, 000--000}


\begintopmatter

\title{Triaxial Haloes and Particle Dark Matter Detection}

\author{N.W.\ Evans$^1$, C.M.\ Carollo$^{2}$ \& P.T.\ de Zeeuw$^3$} 

\vskip0.15truecm
\affiliation{$^1$Theoretical Physics, Department of Physics, 1 Keble Road,
                 Oxford, OX1 3NP}

\vskip0.15truecm

\affiliation{$^2$Department of Astronomy, Columbia University,
		538 W. 120th Street, New York, NY 10027, USA}

\vskip0.15truecm

\affiliation{$^3$Sterrewacht Leiden, Postbus 9513, 2300 RA Leiden, 
                 The Netherlands}

\shortauthor{N.W.\ Evans, C.M.\ Carollo, P.T.\ de Zeeuw} 

\shorttitle{Triaxial Dark Haloes} 


\abstract{This paper presents the properties of a family of
scale-free triaxial haloes.  We adduce arguments to suggest that the
velocity ellipsoids of such models are aligned in conical coordinates.
We provide an algorithm to find the set of conically aligned velocity
second moments that support a given density against the gravity field
of the halo.  The case of the logarithmic ellipsoidal model -- {\it
the simplest triaxial generalisation of the familiar isothermal
sphere} -- is examined in detail. The velocity dispersions required to
hold up the self-consistent model are analytic.  The velocity
distribution of the dark matter can be approximated as a triaxial
Gaussian with semiaxes equal to the velocity dispersions.

There are roughly twenty experiments worldwide that are searching for
evidence of scarce interactions between weakly-interacting
massive-particle dark matter (WIMPs) and detector nuclei.  The annual
modulation signal, caused by the Earth's rotation around the Sun, is a
crucial discriminant between WIMP events and the background.  The
greatest rate is in June, the least in December.  We compute the
differential detection rate for energy deposited by the rare
WIMP-nucleus interactions in our logarithmic ellipsoidal halo models.
Triaxiality and velocity anisotropy change the total rate by up to
$\sim 40 \%$, and have a substantial effect on the amplitude of the
annual modulation signal. The overall rate is greatest, but the
amplitude of the modulation is weakest, in our radially anisotropic
halo models.  Even the sign of the signal can be changed. Restricting
attention to low energy events, the models predict that the maximum
rate occurs in December, and not in June.}
 

\maketitle  


\section{Introduction}

\noindent
The construction of velocity distributions for triaxial haloes is a
hard problem. Jeans' theorem guarantees that the distribution function
depends only on the isolating integrals of motion. Generally, motion
in a triaxial potential admits only one exact integral of motion, the
energy or the Jacobi constant for the case of figure rotation. As
Schwarzschild (1981) has articulated, it seems that self-consistent
triaxial equilibrium configurations can exist only if the potential
has additionally two effective, non-classical integrals.  To date, the
most general and successful method for building triaxial models has
been the numerical combination of orbit densities, often called
Schwarzschild's (1979, 1982) method.  Recent applications and
extensions of the method to triaxial modelling are given by Merritt \&
Fridman (1996), Zhao (1996) and H\"afner et al. (2000).

Given the difficulty of the task, it is often the case that only the
lowest order velocity moments of the distribution function are
calculated via the Jeans equations (sometimes called the stellar
hydrodynamical equations). This is because the moments are easier to
obtain and are related directly to observable properties. Many
solutions of the Jeans equations have been derived for spherical and
axisymmetric systems (e.g., Binney \& Mamon 1982; Bacon 1985; Fillmore
1986; Amendt \& Cuddeford 1991). Even though it is not always evident
whether such solutions correspond to dynamical models with positive
definite distribution functions, these studies have been useful. For
example, starting configurations for $N$-body experiments can be
generated by assuming that the velocity distributions are Gaussians
with semi-axes as prescribed by the Jeans solutions.  Of course, this
is only an approximation, but it is an excellent one and in routine
use (e.g., Barnes 1994).

The simplest model of all for a dark halo is the isothermal
sphere. This is commonly used to estimate the rates in both
microlensing and non-baryonic dark matter detection experiments (e.g.,
Paczy\'nski 1986; Jungman, Kamionkowski \& Griest 1996; Lewin \& Smith
1996). There is no reason whatsoever why dark matter haloes should be
spherical, and there is ample evidence from external galaxies that
haloes are generically flattened (e.g., Sackett et al. 1994; Olling
1995, 1996).  In this paper, we present a number of properties of one
of the simplest triaxial halo models -- the logarithmic ellipsoidal
potential (e.g., Binney \& Tremaine 1987; de Zeeuw \& Pfenniger 1988;
Miralda-Escud\'e \& Schwarzschild 1989).  This is the natural triaxial
generalisation of the singular isothermal sphere.  Self-consistent
distributions of velocities for this model are not known, but here we
provide simple and analytic solutions of the Jeans equations. These
solutions are aligned in conical coordinates and provide the velocity
second moments required to support the triaxial dark halo against
gravity. 

One possibility is that galaxy haloes are composed of
weakly-interacting massive-particle (WIMP) dark matter.  Direct
detection experiments measure the energy deposited by the rare
interactions between the WIMP and the detector nucleus. There are now
about 20 direct detection experiments running or in preparation
worldwide (e.g., articles in Klapdor-Kleingrothaus \& Ramachers 1997
and Spooner 1997). One of the main difficulties in the experiments is
how to distinguish between the recoil events caused by WIMPs and those
caused by radioactivity in the surroundings and by cosmic rays. One
suggestion is that the annual modulation in the WIMP signal, caused by
the motion of the Earth around the Sun, may be a powerful diagnostic
(Freese, Frieman \& Gould 1985). In fact, one of the experimental
groups (DAMA) has very recently claimed detection of this modulation
(Bernabei et al. 1999a,b), though the validity of this claim is still
a matter of fierce debate (e.g., Gerbier et al. 1997, 1999; Abusaiadi
et al. 2000). This paper investigates the dependence of this
differential rate and the annual modulation on the triaxial shape and
velocity distribution of the dark matter halo.

The paper is arranged as follows.  Section 2 introduces the scale-free
halo models under scrutiny, while Section 3 derives and solves the
Jeans equations for the velocity second moments under the assumption
of conical alignment. A number of reasons are given in support of our
assumption as to the orientation of the velocity dispersion tensor.
The properties of the logarithmic ellipsoidal halo models -- the
intrinsic and projected shapes and velocities -- are found in Section
4.  Finally, Section 5 considers the application to direct detection
experiments in some detail.

\beginfigure{1}
\centerline{\psfig{figure=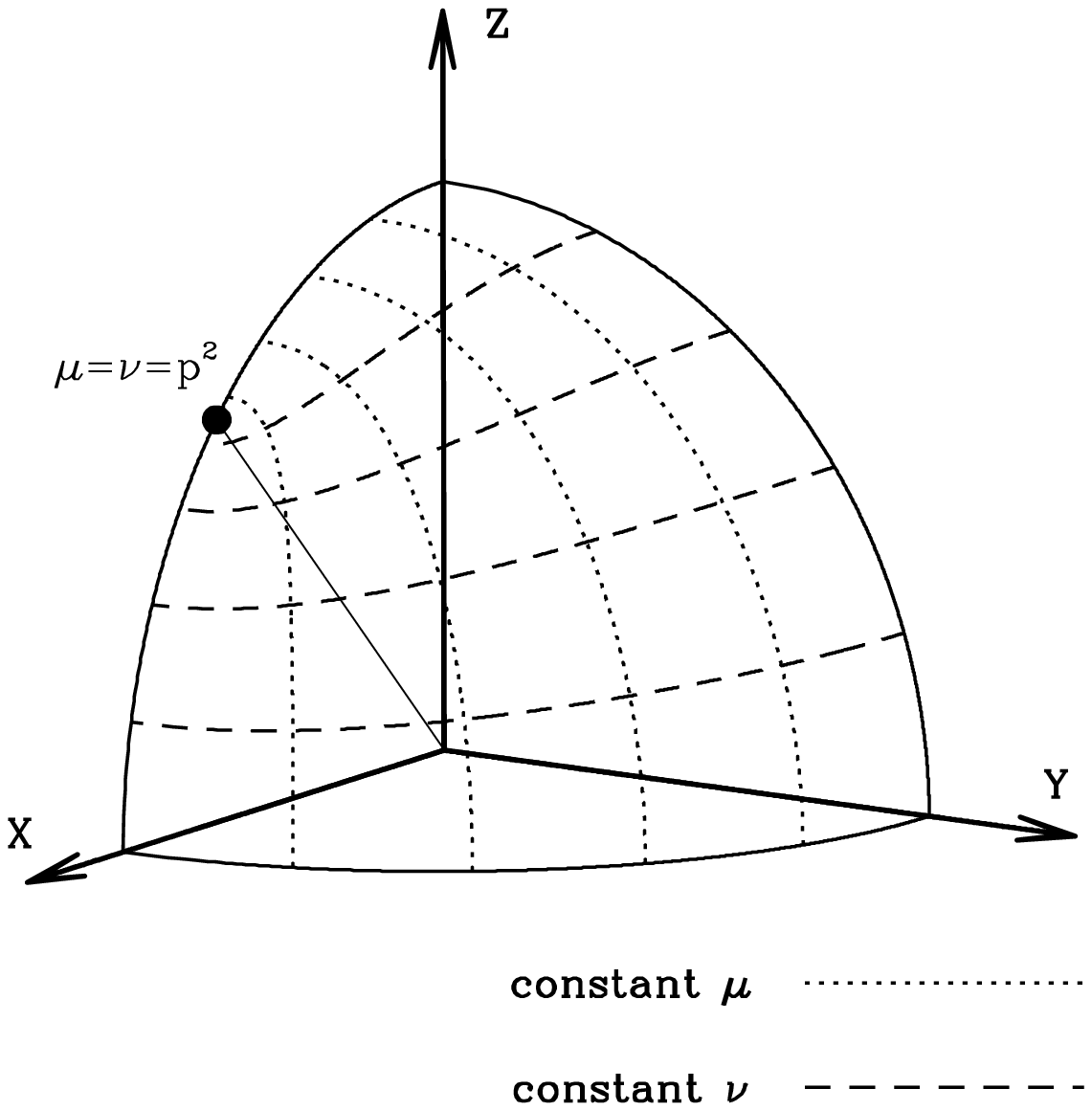,height=8.0truecm}}
\smallskip\noindent
\caption{{\bf Figure 1.} The coordinate curves ($\mu, \nu$) on the
surface of the sphere. If the radius of the sphere is $r$, then the
set ($r,\mu,\nu$) provides a triply-orthogonal set of coordinate
surfaces. The filled dot indicates the point $\mu=\nu=p^2$; it lies in
the $(x, z)$-plane, at $x= r \sqrt{T}, z=r \sqrt{1-T}$, with
$T=(1-p^2)/(1-q^2)$ the triaxiality parameter.}
\endfigure

\eqnumber =1
\def\chaphead{\hbox{2.}}

\section{Triaxial Halo Models} 

\noindent
One of the most widely-used axisymmetric halo models is the
logarithmic scale-free potential investigated by Toomre in the 1970s
(see Toomre 1982) and subsequently studied by others (e.g., Richstone
1980; Evans 1993). Its equipotential surfaces are spheroids.  The
analogous scale-free triaxial halo models have equipotentials that are
triaxial ellipsoids with axis ratios $p$ and $q$ (e.g., Binney 1981;
de Zeeuw \& Pfenniger 1988).  We shall call them {\it the logarithmic
ellipsoidal models}.  In Cartesian coordinates, the potential-density
pair is given by:
\eqnam{\powerpot}
$$\Phi(x, y, z)
        = \fr12 \vc^2 \ln (x^2 +  y^2p^{-2} + z^2q^{-2}) \eqno\new$$
and
\eqnam{\powerdens}
$$\rho(x, y, z) = {\vc^2\over 4 \pi G} {A x^2 + B y^2p^{-2} +
                 Cz^2q^{-2}\over (x^2 + y^2 p^{-2} + z^2 q^{-2})^{2}},
                                                              \eqno\new$$
with
$$\eqalign{A &= p^{-2}+ q^{-2} -1,\qquad B = 1 + q^{-2} - p^{-2},\cr 
C &= 1 + p^{-2} - q^{-2}.\cr}\eqno\new$$
The rotation curve is completely flat with amplitude $\vc$. This
potential therefore describes a triaxial halo with a flat rotation
curve.  Without loss of generality, we require $q^2 \le p^2 \le 1$.
The spherical limit ($p=q=1$) is of course the familiar singular
isothermal sphere. [If desired, models with a finite core-radius
$R_{\rm c}$ can be obtained by adding $R_{\rm c}^2$ to the term in
parentheses in the expression for the potential.  These models are
recognized as the triaxial generalization of the axisymmetric
power-law galaxies (Evans 1993, 1994)].

As will become apparent, there is much advantage in working in conical
coordinates $(r, \mu, \nu)$ defined as follows. The first coordinate
$r$ is the distance to the origin, so that $r^2 = x^2 + y^2 + z^2$,
where $(x, y, z)$ are the standard Cartesian coordinates. The
variables $\mu$ and $\nu$ are angles. They are the solutions for
$\tau$ of
\eqnam{\munuconical}
$${x^2 \over \tau-1} + {y^2 \over \tau-p^2} 
   + {z^2 \over \tau-q^2} = 0,                                  \eqno\new$$
with $p$ and $q$ constants that satisfy the condition $0 \leq q \leq p
\leq 1$. This is a quadratic equation for $\tau$ with two real roots,
which we order so that $q^2 \leq \nu \leq p^2 \leq \mu \leq 1$. Some
properties of these coordinates, including the relation with the
standard spherical coordinates $(r, \theta, \phi)$, are given in
Appendix A (see also Morse \& Feschbach 1953).  

Figure 1 illustrates the $(\mu,\nu)$ coordinate curves on
the sphere.  The angular coordinates $\mu$ and $\nu$ have a center at
$\mu=\nu=p^2$, which lies at $\phi=0$ and $\theta=\theta_f$, with
$$\sin^2\theta_f = T \equiv {1-p^2 \over 1-q^2},                \eqno\new$$
where $T$ is the triaxiality parameter (e.g., Franx, Illingworth \& de
Zeeuw 1991). While each choice of $p$ and $q$ defines a conical
coordinate system, all choices which lead to the same value of $T$
correspond to the same set of coordinates.

The logarithmic ellipsoidal models are the simplest members of the more
general family of {\it logarithmic scale-free models} which have
potentials of the form:
\eqnam{\genpot}
$$\Phi(r, \mu, \nu) 
        = \fr12 v_{\rm c}^2 \ln r^2 g(\mu,\nu).                \eqno\new$$
The function $g(\mu, \nu)$ is arbitrary, and describes the angular
dependence of the potential. We write the density distribution in our
models in the general form:
\eqnam{\genden}
$$\rho(r, \mu, \nu) = {h(\mu,\nu) \over r^\gammamod}.             \eqno\new$$
Here $h(\mu,\nu)$ is another arbitrary function that describes the
angular dependence of the density. When $\rho$ is the self-consistent
density, then $\gammamod=2$ and the functions $h(\mu,\nu)$ and
$g(\mu,\nu)$ are related by the Poisson equation $4\pi G \rho =
\nabla^2 \Phi$. This case is useful for the description of the
dominant, dark matter component of haloes. When $\gammamod > 2$, the
density law describes a tracer population (such as Population II stars
or globular clusters) that resides in the halo but falls off faster
than the dark matter.  The Laplacian in conical coordinates is given
in equation (A6). We find:
\eqnam{\genh}
$$\eqalign{
h(\mu,& \nu)  = {1 \over (\mu-\nu) g^2} 
           \Bigl\{ (\mu-\nu)g^2  + 
           g \Bigl[ w'(\nu) {\partial g \over \partial \nu} 
                   -w'(\mu) {\partial g \over \partial \mu} \Bigl] \cr
           & + 2w(\nu) \Bigl[g {\partial^2 g \over \partial \nu^2}  -
                     \bigl({\partial g \over \partial \nu}\bigr)^2 \Bigr] 
           - 2w(\mu) \Bigl[g {\partial^2 g \over \partial \mu^2} - 
                     \bigl({\partial g \over \partial \mu}\bigr)^2 \Bigr]
                                                  \Bigr\},\cr}$$
where $w(\tau) = (\tau -1 )(\tau -p^2) (\tau -q^2)$, and the prime
indicates the derivative with respect to the argument. We have taken
$4\pi G = 1$ and $\vc = 1$.

The logarithmic ellipsoidal models are just a special case with a
simple form of $g(\mu,\nu)$:
\eqnam{\binneygtheta}
$$g(\mu,\nu) = {\mu \nu \over p^2 q^2}.               \eqno\new$$
The associated self-consistent density (\powerdens) can be written in
the form (\genden) with $\gammamod =2$ and with:
\eqnam{\binneyhtheta}
$$h(\mu,\nu ) = {1 \over \mu^2 \nu^2} 
                \bigl[ 2 p^2 q^2 (\mu +\nu) 
            - (p^2 + q^2 + p^2 q^2) \mu\nu \bigr].     \eqno\new$$
In what follows, we derive Jeans solutions for general scale-free
densities in logarithmic scale-free potentials in the next Section,
but return to the self-consistent logarithmic ellipsoidal models
as our illustrative examples in Sections 4 and 5.

\eqnumber =1
\def\chaphead{\hbox{3.}}

\section{The Jeans Equations} 

\subsection{Conical Coordinates} 

\noindent
As Binney \& Tremaine (1987) point out, there is a fundamental defect
with the Jeans equations -- namely, that there is no equation of state
relating the components of the velocity dispersion to the density.
Any method of solving the Jeans equations therefore requires
assumptions regarding either the shape or the orientation of the
velocity ellipsoid. And, of course, the usefulness of such solutions
depends on whether the assumptions are physically motivated.  

We believe that Jeans solutions aligned roughly with conical
coordinates do correspond to physical distribution functions for
triaxial scale-free models. There are three reasons for this. First,
the motion in the separable models is described naturally in
ellipsoidal coordinates, and many positive definite distribution
functions exist (Statler 1987; Hunter \& de Zeeuw 1992). At large
radii the density in separable models becomes scale-free, and the
ellipsoidal coordinates reduce to conical coordinates. Hence, the
velocity ellipsoid is conically aligned. Second, in the Jeans
solutions of axisymmetric scale-free models (see de Zeeuw, Evans
\& Schwarzschild 1996; Evans, H\"afner \& de Zeeuw 1997), approximate
distribution functions are constructed for solutions close to
alignment in spherical polar coordinates.  Conical coordinates appear
to be the most natural generalisation of this result to triaxial
scale-free models. Finally, Amendt \& Cuddeford (1991) have argued
that Jeans solutions for which the kurtosis vanishes can be regarded
as physical. For triaxial scale-free models, the vanishing of the
kurtosis implies alignment in conical coordinates. These arguments are
suggestive, rather than rigorous, but they seem worth pursuing to us.

Let us consider a triaxial halo with a gravitational potential
$\Phi=\Phi(r, \mu, \nu)$ and with a phase-space distribution function
$f = f(r, \mu, \nu, v_r, v_\mu, v_\nu)$. In equilibrium, $f$ must
satisfy the collisionless Boltzmann equation (Binney \& Tremaine 1987,
\S 4.1) which, in conical coordinates, becomes:
\eqnam{\cbe1}
$$\eqalign{
0 &=  r v_r {\partial f \over \partial r} 
  + \Bigl( v_\mu^2 + v_\nu^2 - r {\partial \Phi \over \partial r} \Bigr)      
  {\partial f \over \partial v_r} 
  - v_r v_\mu {\partial f \over \partial v_\mu}
  - v_r v_\nu {\partial f \over \partial v_\nu} \cr
  &\quad + {1 \over Q} \Bigl[ v_\mu {\partial f \over \partial \mu} +
              \Bigl({v_\nu^2 \over 2 (\mu - \nu)} - {\partial \Phi \over
                                                     \partial \mu} \Bigr)
              {\partial f \over \partial v_\mu}
              -{v_\mu v_\nu \over 2 (\mu - \nu)}
               {\partial f \over \partial v_\nu}\Bigr] \cr
   &\quad    + {1 \over R} \Bigl[ v_\nu {\partial f \over \partial \nu} +
              \Bigl({v_\mu^2 \over 2 (\nu - \mu)} - {\partial \Phi \over
                                                     \partial \nu} \Bigr)
              {\partial f \over \partial v_\nu}
              -{v_\nu v_\mu \over 2 (\nu - \mu)}
               {\partial f \over \partial v_\mu}\Bigr],     \cr}$$
where $Q$ and $R$ are the metric coefficients given in equation (A5).
Multiplication by $v_r$, $v_\mu$, and $v_\nu$, respectively, and
subsequent integration over velocity space, gives the Jeans
equations. These relate the mass density $\rho$ and the elements of
the stress tensor $\rho\langle v_i v_j \rangle $ (with $i, j$ equal to
$r, \mu, \nu$) to the forces.  We obtain
\eqnam{\jeans}
$$\eqalign{&{\partial \rho \langle v_r^2 \rangle \over \partial r} 
  + {\rho \over r} \Bigl[ 2 \langle v_r^2 \rangle 
          - \langle v_\mu^2 \rangle - \langle v_\nu^2 \rangle \Bigr] 
  + {1 \over r Q} 
   \Bigl[ {\partial \rho \langle v_r v_\mu \rangle \over \partial \mu}
  + \cr & \qquad { \rho \langle v_r v_\mu \rangle \over 2 (\mu -
  \nu)} \Bigr] + {1 \over r R} 
   \Bigl[ {\partial \rho \langle v_r v_\nu \rangle \over \partial \nu}
  + { \rho \langle v_r v_\nu \rangle \over 2 (\nu-\mu)} \Bigr]
 = - \rho {\partial\Phi \over \partial r},\cr}\eqno\first$$
$$\eqalign{&{\partial \rho \langle v_r v_\mu \rangle \over \partial r} 
 + 3 {\rho \over r} \langle v_r v_\mu \rangle 
 + {1 \over r R } 
  \Bigl[ {\partial \rho \langle v_\mu v_\nu \rangle \over \partial \nu}
  + {\rho \langle v_\mu v_\nu \rangle \over (\nu - \mu)} \Bigr]\cr 
  &\qquad\quad + {1 \over r Q } \Bigl[ {\partial \rho 
   \langle v_\mu^2 \rangle \over \partial \mu}
 + {\rho (\langle v_\mu^2 \rangle - \langle v_\nu^2 \rangle)  
       \over 2 (\mu - \nu)} \Bigr] 
   = -{\rho \over r Q} {\partial\Phi \over \partial \mu},\cr}
                                       \eqno\last{b}$$
$$\eqalign{&{\partial \rho \langle v_r v_\nu \rangle \over \partial r} 
+ 3 {\rho \over r} \langle v_r v_\nu \rangle 
             + {1 \over r Q } 
     \Bigl[ {\partial \rho \langle v_\nu v_\mu \rangle  \over \partial \mu}
            + {\rho \langle v_\nu v_\mu \rangle \over (\mu - \nu)}
\Bigr] \cr
&\qquad\quad
            + {1 \over r R } \Bigl[ {\partial \rho \langle v_\nu^2 \rangle 
                                    \over \partial \nu} + 
   {\rho (\langle v_\nu^2 \rangle - \langle v_\mu^2 \rangle) 
        \over 2 (\nu -\mu)} \Bigr] 
 = -{\rho \over r R} {\partial\Phi \over \partial \nu }.\cr}\eqno\last{c}$$
These three relations between the six stresses $\rho \langle v_r^2
\rangle$, $\rho \langle v_\mu^2 \rangle$, $\rho \langle v_\nu^2
\rangle$, $\rho \langle v_r v_\mu \rangle$, $\rho \langle v_r v_\nu
\rangle$, and $\rho \langle v_\mu v_\nu \rangle$ must be satisfied at
any point in a triaxial halo model with potential $\Phi$ and mass
density $\rho$. A solution of these equations for given $\rho$ and
$\Phi$ corresponds to a physical equilibrium model only if it is
associated with a distribution function $f(r, \mu, \nu, v_r, v_\mu,
v_\nu) \geq 0$.

These three partial differential equations must be supplied with
boundary conditions. The stresses must all vanish at
infinity. Furthermore, the factor $(\mu-\nu)$ vanishes at the special
point $\mu=\nu=p^2$. In order to avoid singularities in the terms on
the left-hand side, when $\mu = \nu = p^2$ we must have (c.f., Evans
\& Lynden-Bell 1989)
\eqnam{\boundaryconditions}
$$ \langle v_\mu^2 \rangle = \langle v_\nu^2 \rangle, \qquad \langle
   v_\mu v_\nu \rangle = \langle v_r v_\mu \rangle = \langle v_r v_\nu
   \rangle = 0.  \eqno\new$$
At this point, the velocity ellipsoid is isotropic in the angular
direction. However, the radial dispersion $\langle v_r^2
\rangle$ may differ from $\langle v_\mu^2 \rangle = \langle v_\nu^2
\rangle$ here. 

\subsection{The Scale-Free Ansatz} 

The potential (\genpot) and density (\genden) have the desirable
attribute of scale-freeness. Their properties at radius $r' = kr$
follow from those at radius $r$ by a simple magnification, and by a
rescaling of the time variable $t' = kt$. For example, $\rho(kr, \mu,
\nu) = k^{-\gammamod} \rho(r,\mu, \nu)$. We consider distribution
functions that are also scale-free, i.e., that satisfy $f(kr, \mu,
\nu, v_r, v_\mu, v_\nu) = k^{-\gammamod} f(r, \mu, \nu, v_r, v_\mu,
v_\nu)$. The associated stresses then have the following form:
\eqnam{\stressansatz}
$$\eqalign{
\rho \langle v_r^2 \rangle       &= {F_1(\mu,\nu) \over
r^{\gammamod}}, \qquad\qquad
\rho \langle v_\mu v_\nu \rangle = {F_4(\mu,\nu) \over r^{\gammamod}}, \cr
\rho \langle v_\mu^2 \rangle     &= {F_2(\mu,\nu) \over
r^{\gammamod}},\qquad\qquad 
\rho \langle v_r v_\mu \rangle   = {F_5(\mu,\nu) \over r^{\gammamod}}, \cr
\rho \langle v_\nu^2 \rangle     &= {F_3(\mu,\nu) \over r^{\gammamod}},\qquad\qquad
\rho \langle v_r v_\nu \rangle   = {F_6(\mu,\nu) \over r^{\gammamod}}, \cr}
                                              \eqno\new$$
where $F_1$, $F_2$, $F_3$, $F_4$, $F_5$ and $F_6$ are functions of
$\mu, \nu$.  A necessary - but not sufficient - condition for the
stresses (\stressansatz) to correspond to a physical equilibrium model
is that $F_1, F_2, F_3 \geq 0$, since they give the velocity average
of the non-negative quantities $v_r^2$, $v_\mu^2$, and $v_\nu^2$.
$F_4$, $F_5$ and $F_6$ may be negative, but we always require that the
eigenvalues of the stress tensor are non-negative. Even then, not all
such solutions correspond to positive distribution functions.

Substitution of the forms (\stressansatz) into the Jeans equations leads to a
system of three partial differential equations for the six variables
$F_{1,\dots,6}$. We find :
\eqnam{\scalejeans}
$$\eqalign{(\gammamod-2) F_1 + F_2 + F_3 & = h + 
       {1 \over Q} \Bigl[ {\partial F_5 \over \partial \mu}  
                          + {F_5 \over 2 (\mu-\nu)} \Bigr] \cr
   &  + {1 \over R} \Bigl[ {\partial F_6 \over \partial \nu}  
                          + {F_6 \over 2 (\nu-\mu)} \Bigr]}
                           \eqno\first$$
$$\null\qquad 
\eqalign{\Bigl[ {\partial F_2 \over \partial \mu}  
                     +  {F_2 - F_3 \over 2 (\mu- \nu)} \Bigr] 
  &= - {h \over 2g} {\partial g \over \partial\mu} 
    + Q(\gammamod-3) F_5 \cr &
   - {Q \over R} \Bigl[ {\partial F_4 \over \partial\nu}       
   + {F_4 \over (\nu-\mu)} \Bigr],}\eqno\last{b}$$
$$\null\qquad
\eqalign{\Bigl[ {\partial F_3 \over \partial \nu} 
                     + {F_3 - F_2 \over 2 (\nu - \mu)} \Bigr] 
   &= - {h \over 2g} {\partial g \over \partial\nu} 
     + R(\gammamod-3) F_6 \cr & - 
     {R \over Q} \Bigl[ {\partial F_4 \over \partial\mu} 
                    + {F_4 \over (\mu- \nu)} \Bigr]} \eqno\last{c}$$
We shall find it convenient to refer to the right-hand
sides of (\scalejeans a), (\scalejeans b) and (\scalejeans c) as
$K_1(\mu,\nu), K_2(\mu,\nu)$ and $K_3(\mu,\nu)$, respectively.  We are
free to pick three of the six functions $F_{1,\dots,6}$ arbitrarily
and solve (\scalejeans) for the other three.

\subsection{Solution for Conical Alignment}

Suppose we pick $F_4, F_5$ and $F_6$ and solve for $F_1, F_2$ and
$F_3$.  This choice is motivated by the fact that it is then easy to
consider the special case $F_{4,5,6}\equiv0$, in which the velocity
ellipsoid is everywhere aligned exactly along the conical coordinate
system. We are particularly interested in this alignment, as we have
argued that it is likely to correspond to physical distribution
functions.

For the special case $\gammamod=2$ (the dark halo case), a simple
approach can be used.  In this case $F_1$ drops out of equation
(\scalejeans a). We use the remaining relation to rewrite (\scalejeans
b, c) as follows:
\eqnam{\scaleone}
$$\eqalign{{\partial F_2 \over \partial \mu}  + {F_2 \over  \mu - \nu} =&
    {h \over 2} \Bigl[ {1 \over \mu-\nu} - 
                       {1 \over g} {\partial g \over \partial \mu} \Bigr]
    + {c_2 \over 2(\mu-\nu)},\cr
{\partial F_3 \over \partial \nu}  + {F_3 \over  \nu - \mu} =&
    {h \over 2} \Bigl[ {1 \over \nu-\mu} - 
                       {1 \over g} {\partial g \over \partial \nu} \Bigr]
    + {c_3 \over 2(\mu-\nu)},\cr}                          \eqno\new$$ 
where the functions $c_2(\mu, \nu)$ and $c_3(\mu, \nu)$ are defined as 
\eqnam{\ctermsdef}
$$\eqalign{c_2 =& 2{Q\over R} \Bigl[ F_4+(\nu-\mu){\partial F_4 \over \partial \nu}\Bigr]
       + {1 \over Q} \Bigl[ {\partial F_5 \over \partial \mu}  
                          + {F_5 \over 2 (\mu - \nu)} \Bigr] \cr  
       &\qquad\qquad -2(\mu-\nu)Q F_5 
       + {1 \over R} \Bigl[ {\partial F_6 \over \partial \nu}  
                          + {F_6 \over 2 (\nu - \mu)} \Bigr],\cr}
\eqno\first$$
$$\eqalign{c_3 =& 2{R\over Q} \Bigl[ F_4+(\mu-\nu){\partial F_4 \over \partial \mu}\Bigr]
       + {1 \over Q} \Bigl[ {\partial F_5 \over \partial \mu}  
                          + {F_5 \over 2 (\mu - \nu)} \Bigr]  \cr
       &\qquad\qquad -2(\nu-\mu)R F_6 
       + {1 \over R} \Bigl[ {\partial F_6 \over \partial \nu}  
                       + {F_6 \over 2 (\nu - \mu)} \Bigr].\cr} 
\eqno\last{b}$$
Equations (\scaleone a) and (\scaleone b) can  be integrated
separately, by using $\mu-\nu$ and $\nu-\mu$, respectively, as
integrating factors. The result is
\eqnam{\solveftwofthree}
$$\eqalign{F_2 =&  {1 \over 2 (\mu - \nu)}  
         \Bigl[ G_2(\mu, \nu) 
              + C_2(\mu, \nu) + D(\nu) \Bigr],\cr
F_3 =&  {1 \over 2 (\nu - \mu)} 
         \Bigl[ G_3(\mu, \nu) 
              + C_3(\mu, \nu) + 
                      {\hat D}(\mu) \Bigr],\cr}\eqno\new$$
where, for the moment, the functions $D(\nu)$ and $\hat D(\mu)$ are
arbitrary, and we have introduced the functions
\eqnam{\hcdefs}
$$\eqalign{
G_2 &= \int \limits^\mu \d m \, h(m, \nu) 
       \bigl[1 + {(\nu-m) \over g(m, \nu)} 
                      {\partial g(m,\nu) \over \partial m} \bigr], \cr
G_3 &= \int \limits^\nu \d n \, h(\mu, n) \bigl[1 + {(\mu-n) 
                           \over g(\mu, n)} 
                      {\partial g(\mu,n) \over \partial n} \bigr],\cr}
				                            \eqno\new$$
and
$$C_2 = \int \limits^\mu \d m \, c_2(m, \nu), \qquad\qquad
C_3 = \int \limits^\nu \d n \, c_3(\mu, n),               \eqno\new$$
so that $G_2$ and $G_3$ depend only on the chosen density and
potential.  Equation (\solveftwofthree) gives the general solution for
$F_2$ and $F_3$ when $\gammamod=2$, subject to the requirement on the
functions $\hat D(\mu)$ and $D(\nu)$ imposed by the boundary
conditions (\boundaryconditions), and by equation (\scalejeans
a). This latter requirement can be written as:
\eqnam{\symmetrycondition}
$${\hat D}(\mu) - D(\nu) = 2(\mu-\nu) h +G_3 -G_2 +C_3 -C_2.     \eqno\new$$
The boundary condition (\boundaryconditions) then shows that ${\hat
D}(p^2)=D(p^2)$, so that we can in fact consider $\hat D$ and $D$ the
same function $\Delta(\tau)$, with $\Delta(\mu)={\hat D}(\mu)$ and
$\Delta(\nu) =D(\nu)$. When $F_4$, $F_5$ and $F_6$ are chosen such
that the right-hand side of this equation can be written as a
difference $\Delta(\mu)-\Delta(\nu)$, then the entire Jeans solution
is specified: $F_1$ is also arbitrary (but non-negative), and $F_2$
and $F_3$ are given by (\solveftwofthree). Otherwise, one can specify
only two of the three functions $F_4$, $F_5$, and $F_6$, and attempt
to solve equation (\symmetrycondition) for the third.  For
$F_{4,5,6}\equiv 0$ both $c_2$ and $c_3$ vanish, and the solutions
without cross terms in the stress tensor have $C_2 \equiv C_3 \equiv
0$.

When $\gammamod \neq 2$ (the tracer population case), it is best to
differentiate (\scalejeans b) with respect to $\nu$, differentiate
(\scalejeans c) with respect to $\mu$ and subtract, to obtain the
following equation for the difference $F^*(\mu,\nu) = F_2(\mu,\nu) -
F_3(\mu,\nu)$ :
\eqnam{\fonestareq}
$${\partial^2 F^* \over \partial \mu\partial\nu} 
+\Bigl( {\partial \over \partial \nu} - {\partial \over \partial 
\mu}\Bigr) {F^* \over 2(\mu-\nu)} =  {\partial K_2 \over \partial \nu} -
{\partial K_3 \over \partial \mu}. \eqno\new$$
The right-hand side of (\fonestareq) contains only known
functions. This equation for $F^*(\mu,\nu)$ is identical to an
equation solved by Evans \& Lynden-Bell (1989) by the method of
Green's functions.  If the initial data $F^*(p^2,\nu)$ and $F^*(\mu,
q^2)$ is provided, then $F^*(\mu,\nu)$ is given everywhere as a double
integral over the Green's function (Evans \& Lynden-Bell 1989, section
4). Once we have the difference $F^*(\mu,\nu)$, then it is
straightforward to solve for the individual components:
$$F_2(\mu,\nu) = F_2(p^2,\nu) + \int_{p^2}^\mu \d m\, 
{F^*(m,\nu)\over 2(m - \nu)}.\eqno\new$$
With $F_2(\mu,\nu)$ and $F_3(\mu,\nu)$ now available, $F_1(\mu,\nu)$
is obtained by re-arranging (\scalejeans a). Therefore, the boundary
conditions required are: {\it (i)} the difference $F^*(\mu,\nu) =
F_2(\mu,\nu) -F_3(\mu,\nu)$ on the surface $\mu = p^2$ and on $\nu =
q^2$, together with {\it (ii)} the value of either $F_2(\mu,\nu)$ or
$F_3(\mu,\nu)$ on either $\mu = p^2$ or on $\nu = q^2$.

\beginfigure{2}
\centerline{\psfig{figure=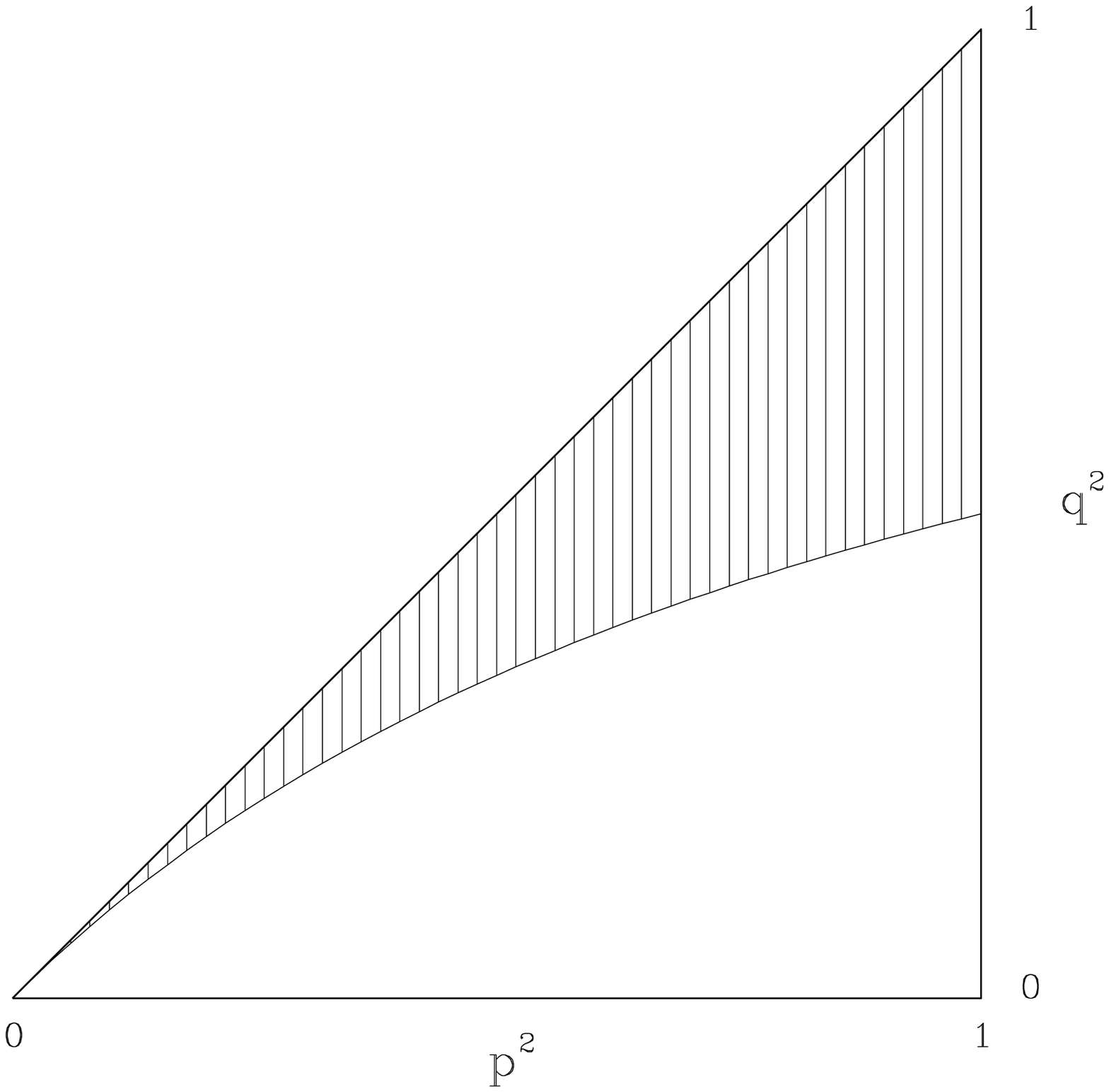,height=5.5truecm,width=5.5truecm}}
%
\caption{{\bf Figure 2.} The logarithmic ellipsoidal models have 
density distributions that are everywhere positive in a hatched region
of the $(p^2, q^2)$-plane. Oblate models have $p^2=1$, while prolate
models have $p^2=q^2$.}
\endfigure
\beginfigure*{3}
\centerline{\psfig{figure=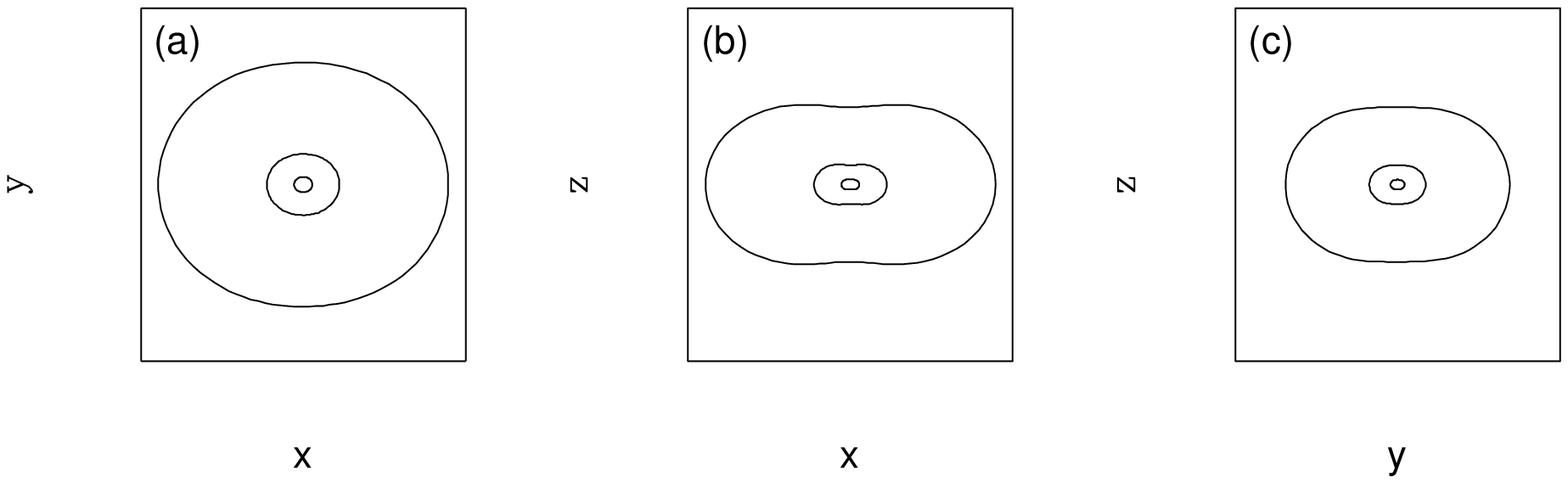,height=6.0truecm}}
\smallskip\noindent
\caption{{\bf Figure 3.} Contour plots of the intrinsic density in the
principal planes for the logarithmic ellipsoidal model with $p=0.9$
and $q=0.8$.}
\endfigure
\beginfigure*{4}
\centerline{\psfig{figure=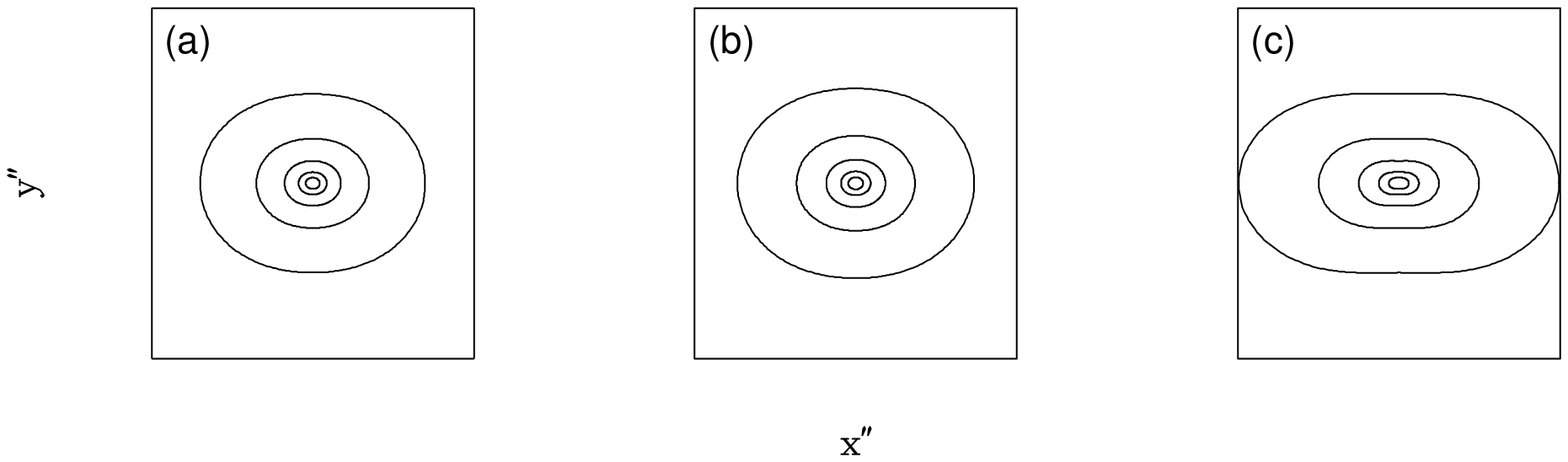,height=6.0truecm}}
\smallskip\noindent
\caption{{\bf Figure 4.} Projected surface density for 
the logarithmic ellipsoidal model with $p=0.9$ and $q=0.8$, and three
different directions of observation, namely a) $\vartheta = 0^\circ$
and $\varphi = 0^\circ$, b) $\vartheta = 30^\circ$ and $\varphi =
45^\circ$, and c) $\vartheta = 90^\circ$ and $\varphi = 90^\circ$. }
\endfigure

\eqnumber =1
\def\chaphead{\hbox{4.}}

\section{The Logarithmic Ellipsoidal Halo}

\subsection{Intrinsic and Projected Shapes}

\noindent
In this section, we return to the logarithmic ellipsoidal models --
partly for their astrophysical importance, partly for their
simplicity.  The density law (\powerdens) is sensibly positive
whenever
$$q^2 + p^2q^2 - p^2 > 0.\eqno\new$$
The physically allowed domain is illustrated as the hatched region in
Figure 2. The oblate models have $p^2=1$, while prolate models have
$p^2=q^2$. Although the equipotential surfaces are ellipsoidal, the
density figures deviate from a pure ellipsoidal shape. Cross-sections
of the intrinsic shape with the three principal planes for the model
with $p =0.9$ and $q=0.8$ are shown in Figure 3. The axis ratios of
the density distribution are
$$\eqalign{\Bigl( {a_2\over a_1} \Bigr)^2 &= {p^2(p^2 q^2 + p^2 - q^2) \over
                                   q^2 + p^2 - p^2q^2},\cr
\Bigl( {a_3 \over a_1 }\Bigr)^2 &=  {q^2(p^2 q^2 + q^2 - p^2) \over
                                   q^2 + p^2 - p^2q^2}.\cr}\eqno\new$$
For our example with $p =0.9$ and $q=0.8$, the density contours have
semiaxes in the ratio $1: 0.774 : 0.489$.  The flattening in the
density is of course greater than the flattening in the potential.

In order to calculate the projected shape, we choose new coordinates
$(x'', y'', z'')$ with the $z''$-axis along the line of sight, and the
$x''$-axis in the $(x, y)$-plane (see e.g., de Zeeuw \& Franx 1989).
The $z$-axis of the triaxial model projects onto the $y''$ axis, and
$x''$ and $y''$ are Cartesian coordinates in the plane of the sky. The
various transformations needed to calculate the projected surface
density are summarised in Appendix B.  The required integration
reduces to an integral given in Appendix C of Evans \& de Zeeuw
(1994).  The projected surface density $\Sigma(x'', y'')$ is analytic
for all viewing angles $(\vartheta, \varphi)$, namely:
$$\Sigma(x'', y'') = \pi pq \, {x''^2 + y''^2 \over 
              (c_1 {x''}^2 - c_2 x''y'' + c_3 {y''}^2)^{3/2}}.
                                                                \eqno\new$$
where 
$$\eqalign{
c_1 &= \sin^2\varphi + p^2 \cos^2\varphi, \cr
c_2 &= 2(1-p^2) \sin\varphi \cos \varphi \cos\vartheta, \cr
c_3 &= \cos^2\varphi \cos^2\vartheta + p^2\sin^2\varphi \cos^2\vartheta 
                                     + q^2\sin^2\vartheta, \cr}
                                                                \eqno\new$$
When the direction of observation lies in one of the three principal
planes $(\vartheta=\pi/2, \varphi=0, \varphi=\pi/2)$, the coefficient
$c_2$ vanishes, and the principal axes of the projected surface
density lie along the $x''$- and $y''$-directions. However, for all
other viewing directions $c_2\not=0$ (for triaxial models with $p\neq 1$),
the minor axis of the projection is misaligned from the projected
short axis of the model, which by definition falls along the
$y''$-axis.  Since our models are scale-free in projection, the
misalignment is independent of radius.  Figure 4 gives contour plots
for the projected surface density of the model with $p=0.9$, $q=0.8$
for three different viewing directions.

Let us introduce polar coordinates $(R', \Theta)$ in the sky plane,
defined by the relations $x'' = -R' \sin \Theta$, $y''= R'
\cos \Theta$. So, the position angle $\Theta$ is defined with respect 
to the $y''$-axis and is measured in the counter-clockwise direction.
The projected surface density is $\Sigma \propto S(\Theta) / R'$.  The
resulting formula can be simplified further by introducing the two
parameters
$$\eqalign{\bmu =& \fr12 (c_1+c_3) + \fr12 \sqrt{(c_3-c_1)^2+c_2^2}, \cr 
  \bnu =& \fr12 (c_1+c_3) - \fr12 \sqrt{(c_3-c_1)^2+c_2^2}.\cr}$$
These are the conical coordinates $(\bmu, \bnu)$ of the direction of
observation defined by the angles $(\vartheta, \varphi)$ (c.f.\
eq. [5.4] of de Zeeuw \& Franx (1989)). We obtain:
\eqnam{\simplifiedsurfacedensity}
$$\Sigma(R',\Theta) = 
          { 2^{3/2}\pi pq \over R' [\bmu+\bnu - (\bmu-\bnu) 
                 \cos(2\Theta-2\Theta_*)]^{3/2}}, 
                                                             \eqno\new$$
where we have defined the misalignment angle $\Theta_*$ by 
\eqnam{\majoraxispositionangle}
$$\tan 2\Theta_* = {c_2 \over c_3-c_1}.\eqno\new$$
If $\Theta_*$ satisfies equation (\majoraxispositionangle), then so
does $\Theta_*+\pi/2$. If we choose $\Theta_*$ to be the root for
which $(c_3-c_1) \cos 2\Theta_* + c_2 \sin 2\Theta_* < 0$, then
$\Theta_*$ is the position angle of the major axis, as measured
counter-clockwise from the $y''$-axis.  Our expression
(\majoraxispositionangle) for the misalignment angle is identical to
that found for the classical ellipsoids, in which the density is
stratified on similar concentric ellipsoids with semi-axes $1:p:q$
(c.f. Stark 1977). Since the potentials (\genpot) are stratified on
such ellipsoids, the major axis of the projected potential (and hence
the projected surface density) must lie along a position angle
$\Theta_*$.  The axis ratio $b'/a'$ of the isophotes, defined by the
condition $\Sigma(b', \Theta_* -\piby2)= \Sigma(a', \Theta_*)$, is
given by
$${b'\over a'}  =  
  \Bigl( {\bnu \over \bmu} \Bigr)^{3/2}.                   \eqno\new$$
If $p =0.9$ and $q=0.8$, the projected surface density has a shape
roughly like E3 if $\vartheta = \varphi = 0^\circ$ and like E5 if
$\vartheta = \varphi = 90^\circ$ (see Figure 4).

\subsection{Velocity Second Moments}
 
\noindent
We now carry through the algorithm of Section 3 and find a simple set
of velocity dispersions that support the logarithmic ellipsoidal
model.  Since $\gammamod=2$, $F_1$ is arbitrary, while $F_4 = F_5 =
F_6$ all vanish for conical alignment.  Upon substitution of the
expressions for $g(\mu,\nu)$ and $h(\mu,\nu)$, the integrals in
(\solveftwofthree) can be carried out and we obtain:
\eqnam{\solveqbin1}
$$\eqalign{F_2(\mu,\nu) =& {1 \over 2  (\mu - \nu)}
\Bigl[(p^2 + q^2 + p^2 q^2 - {2 p^2 q^2 \over \nu}) {1 \over \mu} \cr
&\qquad\qquad - {p^2 q^2 \over \mu^2} + D(\nu)  \Bigr],\cr} \eqno\first$$
\eqnam{\solveqbin2}
$$\eqalign{F_3(\mu,\nu) =& {1 \over 2 (\nu - \mu)}
\Bigl[(p^2 + q^2 + p^2 q^2 - {2 p^2 q^2 \over \mu}) {1 \over \nu} \cr
&\qquad\qquad - {p^2 q^2 \over \nu^2} + \hat D(\mu)  \Bigr].\cr}  \eqno\last{b}$$
The condition $F_2+F_3=h$ can be satisfied by the following choice:
\eqnam{\constinteg}
$$\Delta(\tau) = {3 p^2 q^2 \over \tau^2} - {p^2 + q^2 + p^2 q^2 \over \tau} 
                +A,                                                \eqno\new$$
with $\Delta$ either the function $D$ or the function $\hat D$, $\tau$
respectively either $\nu$ or $\mu$, and $A$ is a constant. The requirement
that $F_2$ and $F_3$ are finite at the point $\mu=\nu=p^2$ means that the 
term in square brackets must vanish there. This means that we must choose 
$A=0$. The expressions for $F_2$ and $F_3$ can then be simplified to
\eqnam{\explsol}
$$\eqalign{
F_2 =& {1 \over 2 \mu^2 \nu^2}  \bigl[(3\mu+\nu)p^2q^2 
                          - (p^2+q^2+p^2q^2)\mu\nu \bigr],\cr
F_3 =& {1 \over 2 \mu^2 \nu^2} \bigl[(\mu+3\nu)p^2q^2 
                          - (p^2+q^2+p^2q^2)\mu\nu \bigr].\cr}\eqno\new$$
Despite the symmetry of $F_2$ and $F_3$ with respect to $\mu$ and
$\nu$, the velocity ellipsoid is anisotropic everywhere, except at the
point $\mu=\nu=p^2$.  Here $F_2=F_3$, while $F_1$ remains arbitrary.

To gain insight into a physically reasonable choice for $F_1$,
let us consider the spherical limit ($p=q=1$). The logarithmic
ellipsoidal model becomes the well-known singular isothermal
sphere. The angular stresses corresponding to (\explsol) become
$$\langle v_\mu^2 \rangle = \langle v_\nu^2 \rangle = {1\over 2},\eqno\new$$
This is recognised as the Jeans solution generated by the constant
anisotropy distribution functions for the isothermal sphere (see
eq. [5.10] in Evans (1994)). The full solutions for the spherically
aligned stresses are
\eqnam{\isostress}
$$\langle v_r^2 \rangle = {1 \over 2 + \gamma},
\qquad \langle v_\theta^2 \rangle = \langle v_\phi^2 \rangle = 
{1 \over 2},\eqno\new$$
where $\gamma$ is a constant anisotropy parameter. In the spherical
limit, $\gamma =0$ gives an isotropic solution, $\gamma \rightarrow
-2$ is the radial orbit model, while $\gamma \rightarrow \infty$ is
the circular orbit model. This suggests a possible choice for $F_1$
as
\eqnam{\wynschoice}
$$F_1 = {g(\mu,\nu) \over 2 + \gamma} = {\mu\nu\over (2+\gamma) p^2q^2}.
\eqno\new$$
Any choice for $F_1$ satisfies the Jeans equations. This choice
(\wynschoice) has the advantage that the spherical limit ($p^2 = q^2 =
1$) certainly corresponds to a positive definite distribution function
for any choice of anisotropy parameter $\gamma > -2$. The inclusion of
the term $g(\mu,\nu)$ in (\wynschoice) additionally ensures that the
axisymmetric limit ($p^2=1$) with $\gamma =0$ has the same stresses as
the two integral distribution function (see eq.\ (2.3) of Evans
(1993)). It is clearly heartening that the checkable limits do
correspond to physical models. Even so, the triaxial models probably
only have physical distribution functions for some range in $\gamma$
near to the isotropic value. Strongly anisotropic models are known to
be afflicted by instabilities, like the radial orbit instability
(e.g., Palmer 1994).  Henceforth, we only consider models in which the
inequality in any two semiaxes of the velocity dispersion tensor is no
greater than $3:1$.

Given our solutions for $F_1, F_2$ and $F_3$, the intrinsic velocity
second moments of the logarithmic ellipsoidal model are found through
(\stressansatz) as;
\eqnam{\mainresult}
$$\eqalign{\langle v_r^2 \rangle =&  {\mu^3\nu^3 \over (2 + \gamma)p^2q^2}
                         {1\over [2(\mu+\nu)p^2q^2 -(p^2+q^2+p^2q^2)
                          \mu\nu]},\cr
\langle v_\mu^2 \rangle =& {1 \over 2}\, {(3\mu+\nu)p^2q^2 
                          - (p^2+q^2+p^2q^2)\mu\nu \over
                             2(\mu+\nu)p^2q^2 - (p^2+q^2+p^2q^2)\mu\nu},\cr
\langle v_\nu^2 \rangle =& {1 \over 2}\, {(\mu+3\nu)p^2q^2 
                          - (p^2+q^2+p^2q^2)\mu\nu \over
                             2(\mu+\nu)p^2q^2 -
                             (p^2+q^2+p^2q^2)\mu\nu}.\cr}
\eqno\new$$
The relationship between conicals ($\mu,\nu$) and familiar Cartesian
coordinates is given in eqs. (A1) and (A2). This furnishes an analytic
and realistic set of stresses that support the logarithmic ellipsoidal
model -- a prototype for a triaxial dark matter halo. This is one of
the main results of the paper.

The logarithmic ellipsoidal model, much like the isothermal sphere
itself, does not obey the virial theorem (e.g., Binney \& Tremaine
1987, Gerhard 1991). In fact, the velocity dispersions (\mainresult)
have a curious property -- the total kinetic energy is not fixed, and
the radial velocity dispersion may be changed independently of the
angular velocity dispersions.  The same property holds for the
isothermal sphere itself, as is evident from (\isostress). It is a
surprising feature, as we expect increases in the radial velocity
dispersion to be balanced by decreases in the angular dispersions.  In
fact, both isothermal spheres and ellipsoids make complete sense when
regarded as the inner parts of finite mass models. For example, the
truncated, flat rotation curve model (Wilkinson \& Evans 1999) has a
density that varies like $r^{-2}$ within a scalelength $a$ so that:
$$\rho(r) = {a^3\over r^2 (r^2 + a^2)^{3/2}},
\quad \Phi(r) = -\log \Bigl[ {\sqrt{r^2 + a^2} + a \over r} \Bigr],
\eqno\new$$
where, as for the logarithmic ellipsoidal model, we are using units
with $\vc = 4\pi G =1$. The radial velocity dispersion is (see
eq. (11) of Wilkinson \& Evans (1999))
\eqnam{\tfmodel}
$$\eqalign{\langle v_r^2 \rangle =& 
\Bigl( {r \over a} \Bigr)^{\gamma+2} { (r^2 +
a^2)^{3/2} \over a^3} \int_0^{\Phi} {\sinh^{5+\gamma} \phi \over
\cosh^3 \phi} d\phi,\cr
\langle v_\phi^2 \rangle =& \Bigl( {r \over a} \Bigr)^{\gamma+2}
{ (2 + \gamma) (r^2 + a^2)^{3/2} \over 2 a^3}
\int_0^{\Phi} {\sinh^{5+\gamma} \phi \over \cosh^3 \phi} d\phi.
\cr}\eqno\new$$
In the inner parts, the rotation curve is flat with unit amplitude and
the model looks like the isothermal sphere.  In the limit $r <<a$, the
crucial point is that the velocity dispersions (\tfmodel) do indeed
reduce to those of the isothermal sphere (\isostress), as a careful
Taylor expansion demonstrates. This is the case for all $\gamma \ge
-2$. The kinetic energy of the inner, isothermal parts of these models
can be increased -- as it can be balanced by a loss in the kinetic
energy of the outer parts.

\beginfigure*{5}
\centerline{\psfig{figure=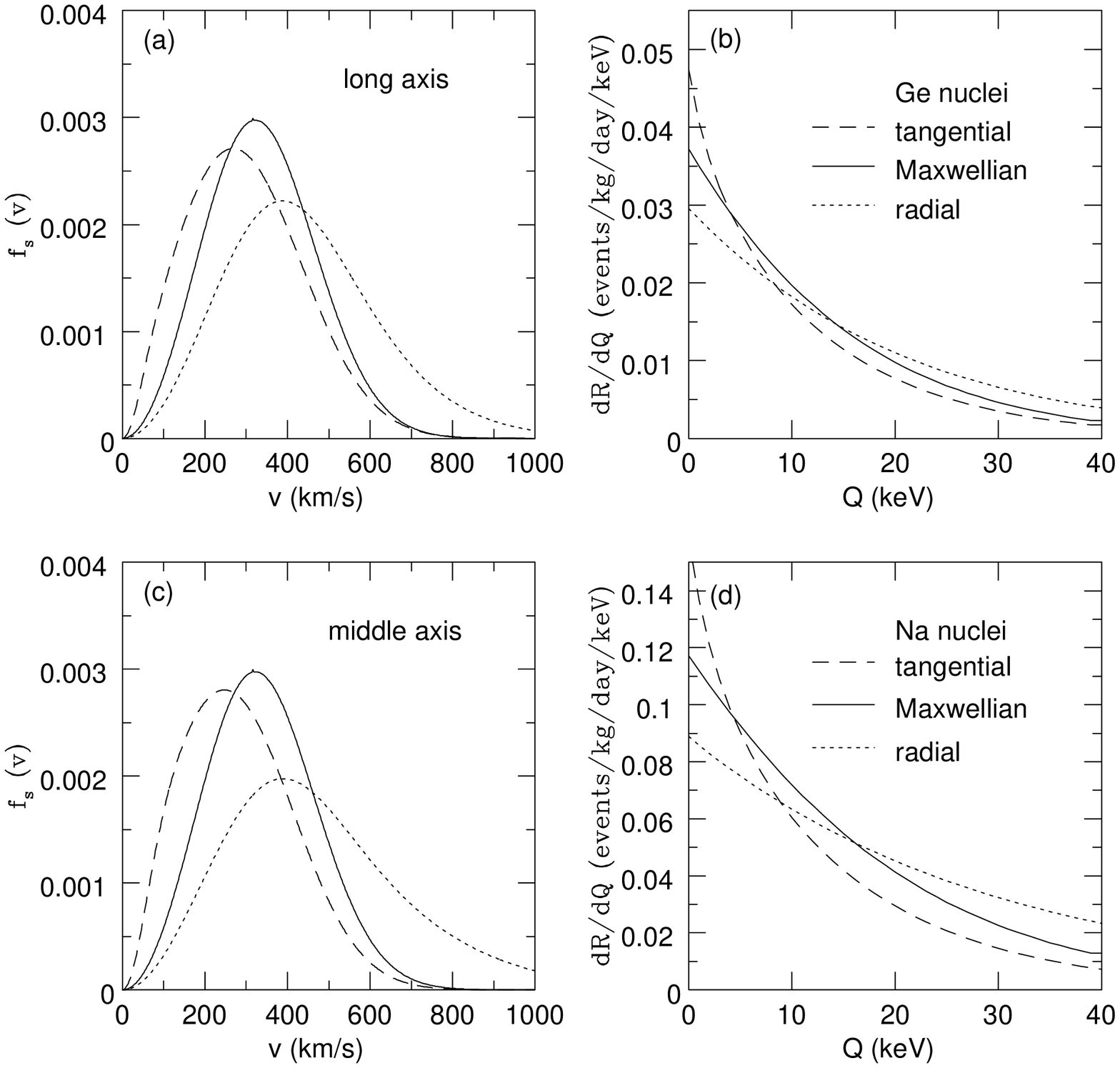,height=14.5truecm}}
\smallskip\noindent
\caption{{\bf Figure 5.} The distribution of speeds and the
differential rate on the major axis (a,b) and the intermediate axis
(c,d) of the logarithmic ellipsoidal model with $p=0.9$ and $q=0.8$.
In each panel, results are given for a radially anisotropic [$\gamma =
-1.78$] solution (dotted line) and tangentially anisotropic [$\gamma
=16$] solution (dashed line), as well as a comparison Maxwellian (full
line).  Panel (b) assumes that the WIMPs are scattering off
${}^{73}$Ge nuclei, while panel (d) assumes that they are scattering
of ${}^{23}$Na nuclei.  The computations are carried out for the date
of June 2nd when the total rate is at a peak.}
\endfigure

\beginfigure*{6}
\centerline{\psfig{figure=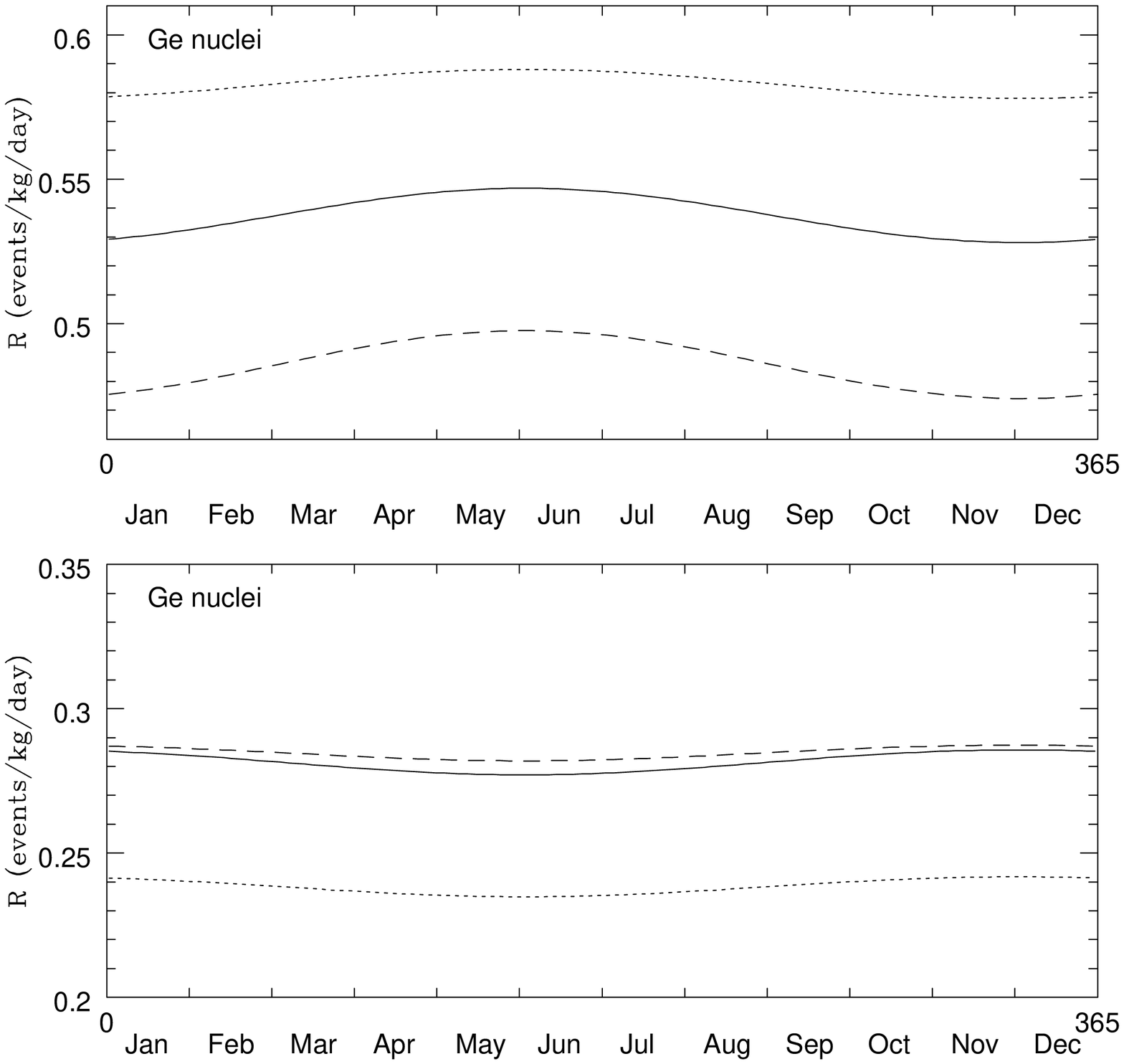,height=12.0truecm}}
\smallskip\noindent
\caption{{\bf Figure 6.} The WIMP annual modulation signal is shown
for a radially anisotropic [$\gamma = -1.78$] model (dotted line), a
tangentially anisotropic [$\gamma = 16$] model (dashed line) and a
comparison Maxwellian (full line). The upper panel gives the variation
in the total rate, the lower panel the variation in the rate of low
energy events ($< 10$ keV). This calculation assumes 40 GeV WIMPs
interacting with ${}^{73}$Ge nuclei. The halo model has $p= 0.9$ and
$q = 0.8$, while the sun is located on the major axis.  }
\endfigure
 
\section{Particle Dark Matter Detection Rates}

\subsection{Preliminaries}

\noindent
The dark matter in galaxy haloes may be composed of weakly-interacting
massive particles (WIMPs), which couple to ordinary matter only
through electroweak-scale interactions. Particle physics has provided
an extensive list of candidates (e.g., Kolb \& Turner 1989), of which
the lightest stable neutral supersymmetric particle (generally the
neutralino) is one of the current favourites (e.g., Jungman et
al. 1996). One promising way of confirming this hypothesis involves
direct detection experiments.  Broadly speaking, the experiments work
by measuring the recoil energy of a nucleus in a low background
laboratory detector which has undergone a collision with a WIMP.  The
aim is to measure the number of events per day per kilogram of
detector material as a function of the recoil energy $Q$. Although
this deposited energy is minute and the WIMP-nucleus interaction is
very rare, there are a number of such experiments in progress around
the world.  These include the UKDMC collaboration operating in Boulby
mine (e.g., Smith et al. 1996), the DAMA collaboration in the Gran
Sasso Laboratory (e.g., Bernabei et al. 1999), which both use NaI
scintillators, and the CDMS experiment located underground at Stanford
University, which uses cryogenic germanium and silicon detectors
(e.g., Gaitskell et al. 1997).

In all these experiments, the detection rate depends on the mass $\mX$
and cross-section $\sigma_0$ of the WIMP, as well as the mass of the
target nucleus $\mN$ in the detector. But, it also depends on the
local dark matter density $\rho_0$ and the speed distribution $\fs
(v)$ of WIMPs in the Galactic halo near the Earth. Calculations have
already been performed using Maxwellian velocity distributions for
singular and cored isothermal spheres, as well as for self-consistent
flattened halo models (e.g., Freese et al. 1985; Jungman et al. 1996;
Kamionkowski \& Kinkhabwala 1998; Belli et al. 1999).  One of our aims
here is to assess the likely uncertainties in the detection rates
caused by halo triaxiality and velocity anisotropy. The formulae for
the calculation of rates in direct detection experiments are
summarised in the review of Jungman et al. (1996). We give here only
the bare details.  The differential rate for WIMP detection is
\eqnam{\diffrate}
$${\d R \over \d Q} = {\sigma_0 \rho_0 \over 2 \mX \mr^2}
                      F^2(Q) \int_\vmin^\infty 
                      {\fs (v) \over v} \d v,\eqno\new$$
where $R$ is the rate, $Q$ is the recoil energy and $\mr = \mN
\mX/(\mN + \mX)$ is the reduced mass, $F(Q)$ is the nuclear form
factor, $\fs (v)$ is the probability distribution of WIMP speeds
relative to the detector, and
$$\vmin  = \Bigl[ { Q\mN \over 2 \mr^2 } \Bigr]^{1/2}.\eqno\new$$
The most commonly-used nuclear form factor is (e.g., Ahlen et al.
1987; Freese et al. 1988; Jungman et al. 1996)
$$F(Q) = \exp\Bigl(-{Q\over 2Q_0}\Bigr), \eqno\new$$
where $Q_0$ is the nuclear coherence energy
$$Q_0 =  {3 \hbar^2\over 2\mN \RN^2}, \eqno\new$$
and $\RN$ is the radius of the target nucleus (in cm)
$$\RN = 10^{-13}\bigl[0.3 + 0.91(\mN/{\rm GeV})^{1/3}\bigr].\eqno\new$$
The total event rate can be found by integrating over all detectable
energies
$$R = \int_\ET^\infty {\d R \over \d Q} \d Q,\eqno\new$$
where $\ET$ is the threshold energy for the detector.

As our benchmark model, we take a WIMP with only scalar interactions
of mass $\mX = 40$ GeV and a cross-section $\sigma_0 = 4 \times
10^{-36} \cm^2$. We consider two kinds of detectors. The first is a
cryogenic detector made of germanium so that $\mN = 68$ GeV. The
second is a scintillation detector made of NaI, for which results are
presented for WIMPs scattering off sodium nuclei ($\mN = 22$ GeV).
Threshold effects are neglected, so that $\ET =0$.  The local halo
density $\rho_0$ is taken as $0.3 \;{\rm GeV}\cm^{-3}$.  These values
are suggested by Jungman et al. (1996) as a standard set. We note that
they are, of course, subject to very substantial uncertainties. In
particular, speculations as to WIMP masses range from $10$ GeV to a
few TeV.  WIMPs may have both scalar and spin-dependent interactions
with the nucleus.  In fact, direct detection experiments are best
suited for scalar-coupled WIMPs, whilst indirect methods (such as
searching for WIMP annihilation products, like neutrinos, in the Sun)
are more powerful for spin-coupled WIMPs (Kamionkowski, Griest,
Jungman \& Sadoulet 1995).  Finally, the uncertainty in the local halo
density is at least a factor of two, and possibly more (e.g., Bahcall,
Schmidt \& Soneira 1983; Turner 1986; Gates, Gyuk \& Turner 1995)

\beginfigure*{7}
\centerline{\psfig{figure=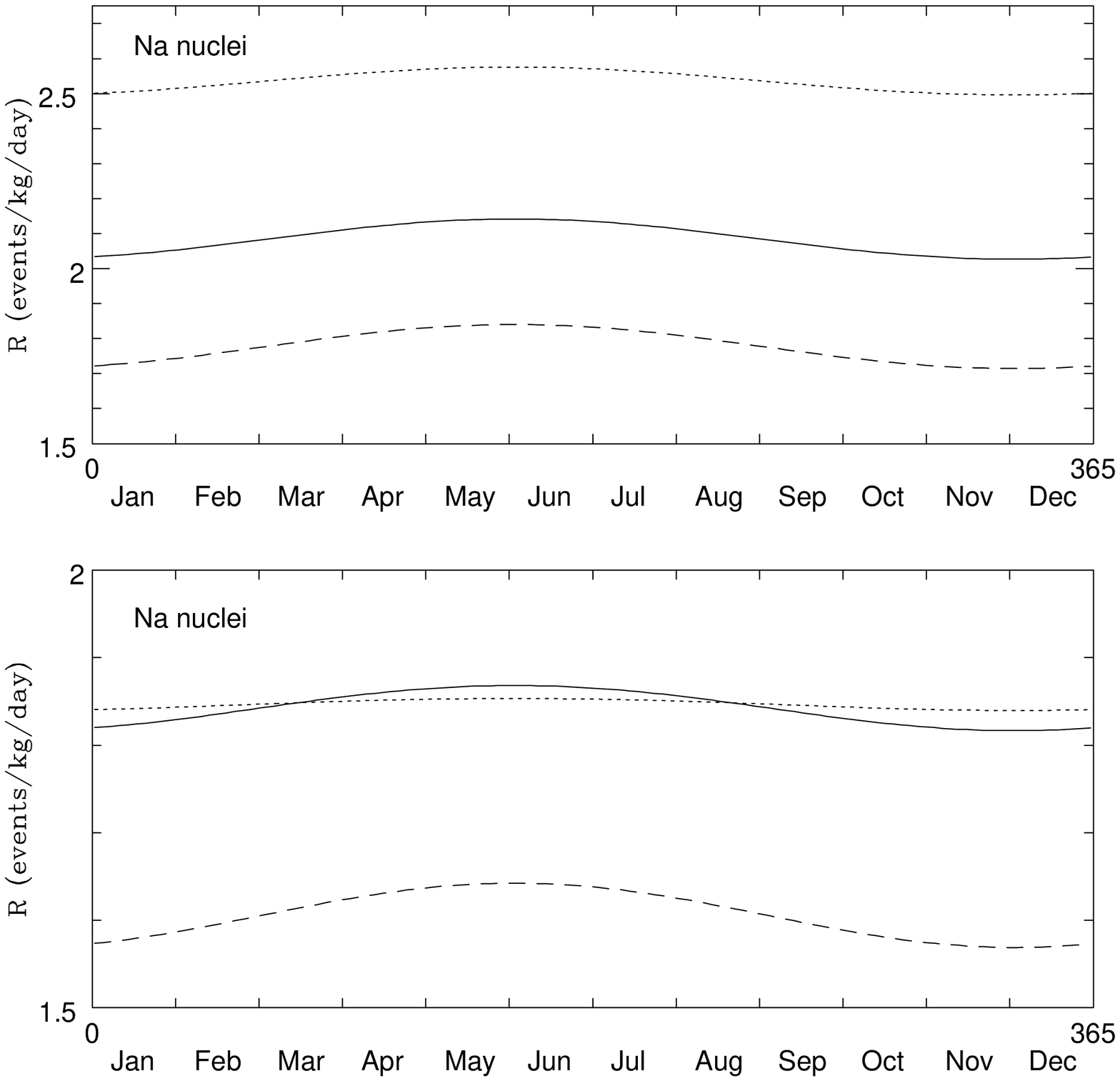,height=12.0truecm}}
\smallskip\noindent
\caption{{\bf Figure 7.} As Figure 6, but the calculation assumes 40
GeV WIMPs scattering off ${}^{23}$Na nuclei. The upper panel shows the
variation in total rate, the lower panel the variation in the rate
of low energy events ($< 10$ keV) assuming a quenching factor of $0.3$.}
\endfigure

\subsection{Triaxiality and Anisotropy}

\noindent
Let us consider two possible locations for the Sun, namely on the long
axis ($x$--axis) and the intermediate axis ($y$-axis) of the triaxial
halo. At these locations, the conical coordinates are locally
equivalent to cylindrical polar coordinates. So, the velocity
distribution in the Earth's rest frame can be approximated as the
triaxial Gaussian
\eqnam{\triGauss}
$$f = {1\over (2\pi)^\threeh \sigR \sigphi \sigz}
                     \exp \bigl[ - {v_R^2 \over 2 \sigR^2}
                    -{(v_\phi + \vearth)^2 \over 2 \sigphi^2}
                    -{v_z^2 \over 2 \sigz^2} \bigr].\eqno\new$$
Here, we assume that the Earth moves with respect to the rest
frame of the Galaxy with a velocity 
$$\vearth = 220 \biggl[1.05 + 0.07 \cos \bigl( {t - \tp \over{\rm 1 yr}} 
\bigr) \biggr]\,\kms,
\eqno\new$$
where $\tp$ corresponds to June 2nd.  This gives an annual modulation
to the WIMP signal, which is of course invaluable in distinguishing it
from background events caused by radioactivity and cosmic rays (Freese
et al. 1988).  Even though many of the experiments operate underground
to cut down the effects from cosmic rays and use high-purity material
to minimise the effects of radioactivity, it is still true that the
background rate is a few events per kilogram per day. This is larger
than that rate expected from WIMP interactions by a factor of $\sim
10$.  The WIMP signal is expected to attain a maximum in early June
and a minimum in early December.  Recently, the DAMA group has claimed
detection of this modulation (Bernabei et al. 1999a,b), although their
interpretation has been contested by others (Gerbier et al. 1999;
Abusaiadi et al. 2000).

On the major axis, the Jeans solution (\mainresult) becomes
$$\eqalign{\sigR^2 =& \langle v_R^2 \rangle = {\vc^2 \over (2 +
           \gamma) (p^{-2} + q^{-2} -1)}, \cr
           \sigphi^2 =& \langle v_\mu^2 \rangle = {\vc^2 (2q^{-2} - 1)
           \over 2 (p^{-2} + q^{-2} -1) },\cr
           \sigz^2 =& \langle v_\nu^2 \rangle = {\vc^2 (2p^{-2} -1)
           \over 2 (p^{-2} + q^{-2} -1)}.\cr} \eqno\new$$
On the intermediate axis, we have
$$\eqalign{\sigR^2 =& \langle v_R^2 \rangle = {\vc^2 p^{-4} \over (2 +
           \gamma) (1  + q^{-2} - p^{-2})}, \cr
           \sigphi^2 =& \langle v_\mu^2 \rangle = {\vc^2 (2q^{-2} - p^{-2})
           \over 2 (1 + q^{-2} - p^{-2}) },\cr
           \sigz^2 =& \langle v_\nu^2 \rangle = {\vc^2 (2 - p^{-2})
           \over 2 (1 + q^{-2} - p^{-2})}.\cr} \eqno\new$$
The distribution of WIMP speeds $\fs (v) $ follows from (\triGauss) as
$$\eqalign{&\fs (v) = {v^2 \over (2\pi)^\threeh \sigR \sigphi \sigz}
\int_0^\pi \d \alpha \sin \alpha \int_0^{2\pi} \d \beta \cr 
& \exp \bigl[ -{(v\cos\alpha + \vearth)^2 \over 2 \sigphi^2}
- {v^2 \sin^2 \alpha \sin^2 \beta \over 2 \sigR^2} 
-{v\sin^2 \alpha \cos^2\beta \over 2 \sigz^2} \bigr].\cr}$$
It is now straightforward to calculate the differential rate
(\diffrate) for any of our triaxial, anisotropic halo models.

Figure 5 shows the distribution of speeds and the differential rate
for the logarithmic ellipsoidal halo with $p = 0.9$ and $q = 0.8$.
The Sun is located on the major axis ($x$-axis) for panels (a) and
(b), the intermediate axis ($y$-axis) for panels (c) and (d). The
curves are drawn for a radially anisotropic and a tangentially
anisotropic velocity distribution that supports the triaxial
figure. Also shown is the standard curve obtained by assuming a
spherical halo with a Maxwellian velocity distribution (c.f., Fig. 22
of Jungman et al. 1996).  Panel (b) shows the case when the WIMPS
interact with germanium detector nuclei, while panel (d) the case of
sodium nuclei.  The effects of triaxiality and velocity anisotropy can
cause the total rate to vary by $\sim 20 \%$ in the case of germanium
and by $\sim 40 \%$ in the case of sodium. The total rate is greatest
for the radial anisotropic velocity distributions, least for the
tangential anisotropic.  In the case displayed in Figure 5, the
radially anisotropic model [$\gamma = -1.78$] has a broader
distribution of speeds. This gives larger incident WIMP velocities,
and so more events at higher energies. The converse is true for the
tangentially anisotropic model [$\gamma = 16$], which generate more
events at lower energies.

More important than the differential rate is the size of the annual
modulation effect. The upper panel of Figure 6 shows the variation in
the total rate during the course of the year for the radially and
tangentially anisotropic halo models. This is for the case of $40$ GeV
WIMPs impinging on germanium nuclei.  Also shown is the comparison
Maxwellian of the isothermal sphere, for which the annual modulation
effect is already small; the peak to peak variation over the course of
the year is just $\sim 0.019$ events/kg/day. For the radially
anisotropic model, this halves to $\sim 0.010$ events/kg/day, while
for the tangentially anisotropic model, this increases slightly to
$\sim 0.024$ events/kg/day. Let us recall that the modulation must be
detected against a background rate that is at least a factor of 10
greater than the underlying WIMP event rate.  Even if the WIMP
velocities are described by a Maxwellian, the detection of the annual
modulation is hard enough. It is disconcerting to find that the
modulation amplitude can halve on moving to radially anisotropic,
flattened models. The lower panel of Figure 6 shows the variation in
the rate of low energy events only ($< 10$ keV).  The amplitude of the
modulation is severely attenuated.  For all the velocity
distributions, the peak to peak variation is very low, at most $0.008$
events/kg/day.  Indeed, even the sign of the modulation has reversed,
and the maximum number of events now occurs in December, not June
(c.f., Hasenbalg 1998).

The DAMA group (Bernabei et al., 1999a,b) have recently extracted
statistical evidence for the annual modulation signal in the low
energy events using nine 9.7 kg NaI detectors. It is therefore
interesting to carry out the calculations for this case as well.
Figure 7 shows the annual variation in rate for WIMPs scattering off
sodium nuclei in NaI detectors. Again, we see that the annual
modulation in the total rate is weakest in the radially anisotropic
model, showing a peak to peak variation of $0.08$ events/kg/day; for
the tangentially anisotropic model, it is $0.12$ events/kg/day.  The
lower panel shows the modulation in the rate of events with {\it
measured} energies less than 10 keV.  Nuclear recoils of energies in
the 1-20 keV range produce little ionisation as they lose energy; most
of the energy goes into phonons. Hence, ionisation or scintillation
detectors are much less efficient than cryogenic detectors.
Accordingly, we have included a ``quenching factor'' of 0.3 to account
for the efficiency of the detector (Spooner et al. 1994; Bernabei et
al. 1996); a measured energy of 10 keV corresponds to a nuclear recoil
energy of $\sim 33$ keV.  We see that the peak to peak variation in
the low energy events (which is where DAMA have claimed evidence of
the signal) is sensitive to the anisotropy.  The lowest variation is
just $0.01$ events/kg/day for the radially anisotropic model; this
rises to $0.07$ events/kg/day for the tangentially anisotropic
model. Reversed modulation (i.e., the maximum occurring in December)
still occurs, but the quenching drives it to still lower energy
ranges. For example, it occurs for events with measured energies less
than 5 keV in the models with both Maxwellian and radially anisotropic
velocity distributions.

\section{Conclusions}

\noindent
In this paper, we have presented a paradigm for the triaxial dark
halo.  The logarithmic ellipsoidal model is the natural generalisation
of the isothermal sphere into the triaxial domain. Many of its
properties are disarmingly simple -- including the potential, mass
density and projected surface density.  This paper has provided
simple, analytic solutions of the Jeans equations for the logarithmic
ellipsoidal model, under the assumption of conical alignment of the
velocity ellipsoid. These solutions give the components of the
velocity dispersion tensor required to hold up the triaxial halo
against gravity.  In this paper, we have adopted the approximation
that the velocity distribution is a triaxial Gaussian with semiaxes
equal to the velocity dispersions as specified by our solutions of the
Jeans equations. While this distribution function does not satisfy
Jeans theorem (e.g., Binney \& Tremaine 1987), it does have the right
stresses or momentum flux to hold the halo up against gravity.  The
logarithmic ellipsoidal models therefore provide simple prototypes for
the density and velocity distributions of triaxial haloes.

Solutions of the Jeans equations contain as a subset the models which
have physical meaning, i.e., which have a non-negative distribution
function. Building distribution functions in the triaxial case is not
easy, since only the energy is known to be an exact integral of motion
in the general case. Merritt \& Fridman (1996), following up an
earlier suggestion by Schwarzschild (1993), have provided numerical
evidence that cusped triaxial systems may undergo slow evolution
towards axisymmetry. This may mean that exact distribution functions
for such models do not exist -- leaving the Jeans equations as one of
the few possible investigative tools for these slowly evolving
systems.

As an application, we have considered direct detection rates of
particle dark matter. Weakly-interacting massive particle (WIMP)
detection experiments typically measure the nuclear recoil energy as
the WIMPs collide with detector nuclei. Predictions for WIMP detection
rates are habitually carried out using the isothermal sphere to model
the dark matter distribution. This paper has provided calculations for
more realistic triaxial dark haloes with anisotropic velocity
distributions. Standard isothermal sphere calculations give estimates
for the total rate that are good to within $\sim 20 \%$ for Ge detectors
and to within $\sim 40 \%$ for NaI detectors. Both the shape of the
differential rate distribution and the size of the annual modulation
effect are sensitive to the velocity distributions. In pessimistic
cases, the modulation amplitude roughly halves on moving to radially
anisotropic and flattened models, rendering detection of this
characteristic WIMP signature even harder.  In optimistic cases,
tangentially anisotropic models can give slightly higher peak to peak
changes than standard isothermal spheres, although the total rate is
smaller.  Assuming a WIMP mass of $40$ GeV and a cross-section
$\sigma_0 = 4 \times 10^{-36} \cm^2$ for Ge nuclei, then the peak to
peak variation in the total rate over the year may then be as low as
$\sim 0.010$ events/kg/day or as high as $\sim 0.024$ events/kg/day,
with the maximum occurring in June and the minimum in December.

If only the low energy events are considered, then the amplitude of
the modulation is smaller still and even the sign of the correlation
can reverse. The greatest number of low energy events can occur in
December, not June. The energy at which this crossover happens depends
on the anisotropy and triaxiality of the halo, as well as the detector
material.  In practice, direct detection experiments monitor the
number of events within energy intervals chosen by the
experimenter. It would be interesting to calculate the optimum energy
ranges for detection of both the modulation in the high energy events
and the reversed modulation in the low energy events. In such
difficult experiments, where the background is typically ten times or
more as strong as the signal, this effect should be exploited to
provide a convincing and unambiguous signature of actual detection.

\section*{Acknowledgments}
\noindent
It is a pleasure to record our gratitude to the late Martin
Schwarzschild for many stimulating conversations.  NWE acknowledges
financial support from the Royal Society. He thanks Subir Sarkar for
many invaluable discussions on particle dark matter and Gilles Gerbier
for pointing out the importance of quenching. CMC and PTdZ thank the
Institute for Advanced Study, Princeton, and the sub-Department of
Theoretical Physics, Oxford, for hospitality.

\eqnumber =1
\def\chaphead{\hbox{A}}
\appendix
\section{Conical Coordinates}

\noindent
Here, we collect some properties of the conical coordinates $(r, \mu,
\nu)$ defined in Section 2.1. The two roots $\mu$ and $\nu$ of
equation (\munuconical) are
\eqnam{\defcan1}
$$\mu,\nu = \fr12 (k_1 \pm \sqrt{k_2}),                            \eqno\new$$
with 
\eqnam{\defcan2}
$$\eqalign{
  r^2 k_1 &= (y^2 + z^2) +p^2 (x^2 + z^2) +q^2 (x^2 + y^2), \cr
  r^4 k_2 &= [(p^2-q^2) x^2 + (1-q^2) y^2 + (1- p^2) z^2]^2 \cr
          &\qquad\qquad\qquad\qquad   + 4 (1-p^2) (1-q^2) y^2 z^2. \cr} 
                                                                  \eqno\new$$
It follows that $\mu+\nu=k_1$ and $4\mu\nu=k_1^2-k_2$ are each simple rational
functions of $x^2$, $y^2$, and $z^2$. The inverse transformations are:
\eqnam{\invcanone}
$$\eqalign{x^2 =& {r^2 (1 - \mu ) (1 - \nu ) \over (1 - p^2) (1-q^2)},
  \qquad y^2 = {r^2 (\mu - p^2) (p^2 - \nu ) \over (1- p^2) (p^2 - q^2)}, \cr
  z^2 =& {r^2 (\mu - q^2) (\nu - q^2) \over (1 -q^2) (p^2 -q^2)}, \cr}
                                                             \eqno\new$$
so that each point $(r,\mu,\nu)$ corresponds to eight points $(\pm x,\pm y,\pm
z)$. The standard spherical coordinates $(r, \theta, \phi)$ are related to the
conical coordinates by
\eqnam{\invcanfour}
$$\eqalign{\cos^2 \theta & = {(\mu - q^2) (\nu - q^2) \over 
                             (1 -q^2) (p^2-q^2)}, \cr
  \tan^2 \phi & = {(\mu - p^2)  (p^2 - \nu ) (1-q^2) 
            \over (1- \mu ) (1 - \nu ) (p^2 - q^2)}.\cr} \eqno\new$$
The metric of the conical coordinate system is given by $d s^2 = d x^2
+ d y^2 + d z^2 = d r^2 + r^2 (Q^2 d \mu^2 + R^2 d \nu^2)$, with 
\eqnam{\metric2}
$$\eqalign{Q^2 =&  {\nu - \mu \over 4 w(\mu)}, 
  \qquad\qquad
  R^2 =  {\mu - \nu \over 4 w(\nu)},\cr  
  w(\tau) =& (\tau -1 )(\tau -p^2) (\tau -q^2).\cr}
                             \eqno\new$$
The element of area $d A$ on the sphere of radius $r$ and the 
Laplacian $\nabla^2$ are given by:
\eqnam{\metric4}
$$\eqalign{d A =& r^2 Q R d \mu d \nu 
      = {r^2 (\mu - \nu) d \mu d \nu 
         \over 4 \sqrt{-w(\mu)} \sqrt{w(\nu)}},\cr
\nabla^2 =& {1 \over r^2} {\partial \over \partial r} r^2 
             {\partial \over \partial r} 
           + {1 \over r^2 QR} {\partial \over \partial \mu} 
            {R \over Q}  {\partial \over \partial \mu}
           + {1 \over r^2 QR} {\partial \over \partial \nu}
            {Q\over R} {\partial \over \partial \nu}.\cr}     \eqno\new$$
The relations between the conical and cartesian velocity components
can be written as: 
$$\eqalign{
v_x &= {x \over r} \bigl[ v_r + {v_\mu \over 2 Q (\mu-1)} +
                         {v_\nu \over 2 R (\nu-1)} \bigr], \cr
v_y &= {y \over r} \bigl[ v_r + {v_\mu \over 2 Q (\mu-p^2)} +
                         {v_\nu \over 2 R (\nu-p^2)} \bigr], \cr 
v_z &= {z \over r} \bigl[ v_r + {v_\mu \over 2 Q (\mu-q^2)} +
                {v_\nu \over 2 R (\nu-q^2)} \bigr], \cr}  \eqno\new$$
or, inverting the relations, by:
$$\eqalign{ 
v_r   &= {x v_x + y v_y + z v_z \over r}, \cr
v_\mu &= {2 Q \over (\nu-\mu) r} \bigl[ (\mu-p^2) (\mu-q^2) x
          v_x \cr &\qquad\qquad 
          + (\mu-1) (\mu-q^2) y v_y + (\mu-1) (\mu-p^2) z v_z \bigr], \cr
v_\nu &= {2 R \over (\mu-\nu) r} \bigl[ (\nu-p^2) (\nu-q^2) x
         v_x \cr &\qquad\qquad
         + (\nu-1) (\nu-q^2) y v_y + (\nu-1) (\nu-p^2) z v_z \bigr]. \cr}                                                                    \eqno\new$$
\eqnumber = 1
\def\chaphead{\hbox{\rm B}}
\section{The Projected Moments of Triaxial Models}
 
\noindent
Here, we derive expressions for the projected moments of a triaxial
model. The projected zeroth moment is the surface density, calculated
in the main body of the paper. Formulae for the projected first and
second velocity moments are also given.

We choose new coordinates $(x'',y'', z'')$ with the $z''$-axis
along the line of sight, and the $x''$-axis in the $(x, y)$-plane
(e.g., de Zeeuw \& Franx 1989). This means that the $z$-axis projects
onto the $y''$-axis. We define the direction of observation by the
viewing angles $(\vartheta, \varphi)$, and write ${\bf r}=(x, y,
z)$. Then the coordinate transformation is
$${\bf r''} = {\bf R_3}{\bf r}, \qquad\qquad 
  {\bf r} = {\bf R_3^{-1}} {\bf r''},                            \eqno\new$$
where 
\eqnam{\transf}
$$\eqalign{{\bf R_3} =& 
\left(\matrix{-\sin\varphi &\cos\varphi &0 \cr 
    -\cos\varphi\cos\vartheta &-\sin\varphi\cos\vartheta &\sin\vartheta\cr
    \cos\varphi\sin\vartheta &\sin\varphi\sin\vartheta&\cos\theta \cr}\right),
\cr
\null & \null \cr
{\bf R_3^{-1}} =&
 \left(\matrix{
   -\sin\varphi &-\cos\varphi\cos\vartheta&\cos\varphi\sin\vartheta\cr 
    \cos\varphi & -\sin\varphi\cos\vartheta & \sin\varphi\sin\vartheta \cr
      0         &\sin\vartheta &\cos\vartheta \cr}\right).\cr}         \eqno\new$$
The projected surface density is $\Sigma$ where: 
$$\Sigma(x'', y'') = \int \d z'' \, \rho({\bf R_3^{-1}} {\bf r''}). \eqno\new$$
The major and minor axis of the projected surface density $\Sigma$ of
a triaxial galaxy generally do not line up with the $x''$- and
$y''$-axes, but lie at a position angle $\Theta_* \not=0$. The
expression for $\Theta_*$ is given in eq. (\majoraxispositionangle).
Let $(x', y')$ be Cartesian coordinates on the plane of the sky
aligned with the principal axes of the projected density. Then;
$${\bf r''} = {\bf R_2}{\bf r'}, \qquad\qquad 
  {\bf r'} = {\bf R_2^{-1}} {\bf r''},                            \eqno\new$$
where 
$$\eqalign{{\bf R_2} =& 
\left(\matrix{\cos\Theta_* &-\sin\Theta_* &0 \cr 
              \sin\Theta_* & \cos\Theta_* &0 \cr
              0               & 0               &1 \cr}\right), \cr
\null&\null\cr
{\bf R_2^{-1}} =&
 \left(\matrix{\cos\Theta_*  & \sin\Theta_* & 0 \cr
               -\sin\Theta_* & \cos\Theta_* & 0 \cr
                0               & 0               & 1 \cr}\right).\cr}
\eqno\new$$

For completeness, let us also outline the calculation of the projected
velocity moments.  By differentiation with respect to time, we find
for the velocity components:
\eqnam{\fundamental}
$${\bf {\dot r}''} = {\bf R_3} {\bf {\dot r}}, \qquad\qquad 
  {\bf {\dot r}} = {\bf R_3^{-1}} {\bf {\dot r}''}.         \eqno\new$$
At a fixed position, we can calculate the intrinsic mean motion by
integrating over all velocities weighted by the normalised velocity
distribution function. Subsequent integration along the line of sight,
weighted with the density distribution, then gives the observed mean
motions:
\eqnam{\meanmotion}
$$\eqalign{\mux =& {1\over \Sigma}\int_{-\infty}^{\infty} \d z''\, 
         \rho \langle {\dot x}'' \rangle,                                   
\quad 
  \muy = {1\over \Sigma}\int_{-\infty}^{\infty} \d z''\,
        \rho \langle {\dot y}'' \rangle,\cr 
\vlos =& \muz = {1\over \Sigma} \int_{-\infty}^{\infty}
          \d z''\, \rho \langle {\dot z}'' \rangle, \cr} \eqno\new$$
The projected second moments follow in a similar way; for example, the
diagonal components of the second-rank tensor are
\eqnam{\dispmotion}
$$\eqalign{\muxx =& {1\over \Sigma} \int_{-\infty}^{\infty} \d z''\, 
        \rho \langle \xppdot^2 \rangle,\cr
\muyy =& {1\over \Sigma} \int_{-\infty}^{\infty} \d z''\,
        \rho \langle \yppdot^2 \rangle,\cr
  \vlossq = \muzz =& {1\over \Sigma}\int_{-\infty}^{\infty} \d z''\,
        \rho \langle \zppdot^2 \rangle.\cr}\eqno\new$$
and similarly for the mixed components. The products of the velocity
components in the integrand can be found by use of equation
(\fundamental).

\bye